\documentclass[aps,prl,reprint,superscriptaddress,showpacs,preprintnumbers,amsmath,amssymb,longbibliography]{revtex4-1}
\usepackage[dvipdfmx]{graphicx}
\usepackage{dcolumn}   
\usepackage{bm}        
\usepackage{amssymb, amsmath, amsfonts}   
\usepackage{lipsum}        
\usepackage{color}
\DeclareGraphicsExtensions{{.pdf}}
\begin{document}
\newcommand{\noter}[1]{{\color{red}{#1}}}
\newcommand{\noteb}[1]{{\color{blue}{#1}}}
\newcommand{\field}{\left( \boldsymbol{r}\right)}
\newcommand{\paren}[1]{\left({#1}\right)}
\newcommand{\vect}[1]{\boldsymbol{#1}}
\newcommand{\uvect}[1]{\tilde{\boldsymbol{#1}}}
\newcommand{\vdot}[1]{\dot{\boldsymbol{#1}}}
\newcommand{\vder}{\boldsymbol{\nabla}}
%
\widetext
%
%
\title{
Instantaneous Normal Modes Reveal Structural Signatures for the
Herschel-Bulkley Rheology in Sheared Glasses
}
\author{Norihiro Oyama}
\email{oyamanorihiro@g.ecc.u-tokyo.ac.jp}
\affiliation{Graduate School of Arts and Sciences, The University of Tokyo, Tokyo 153-8902, Japan}
\affiliation{Mathematics for Advanced Materials-OIL, AIST, Sendai 980-8577, Japan}
\author{Hideyuki Mizuno}
\affiliation{Graduate School of Arts and Sciences, The University of Tokyo, Tokyo 153-8902, Japan}
\author{Atsushi Ikeda}
\affiliation{Graduate School of Arts and Sciences, The University of
Tokyo, Tokyo 153-8902, Japan}
\affiliation{Research Center for Complex Systems Biology, Universal Biology Institute, University of Tokyo, Komaba, Tokyo 153-8902, Japan}

\date{\today}
\begin{abstract}
The Herschel-Bulkley law, a universal constitutive relation, has
been empirically known to be applicable to a vast range of soft
materials, including sheared glasses.
Although the Herschel-Bulkley law has attracted public attention, its structural
origin has remained an open question.
In this letter, by means of atomistic simulation of
binary Lennard-Jones glasses,
we report that the instantaneous normal modes with negative
eigenvalues, or so-called imaginary modes, serve as the structural
signatures for the Herschel-Bulkley rheology in sheared glasses.
\end{abstract}
\maketitle
%
Many soft materials are
empirically known to obey a universal constitutive relation, or the
so-called Herschel-Bulkley (HB) law $\langle\sigma\rangle-\sigma_{\rm
Y}=A\dot{\gamma}^n$~\cite{Herschel1926}, where $\langle\sigma\rangle$
is the steady-state average of the shear stress $\sigma$,
$\sigma_{\rm Y}$ is the yield stress, $\dot{\gamma}$
is the strain rate, $n$ is the HB exponent, and $A$ is a coefficient.
Examples encompass, for instance, foams~\cite{Gilbreth2006},
emulsions~\cite{Becu2006,Dinkgreve2015a}, microgel
suspensions~\cite{Moller2009,Gutowski2012}, soft athermal particles~\cite{Olsson2012,Oyama2019a}, blood~\cite{Gosselin2016},
vegetables, and fruits~\cite{Diamante2015}.
Experiments~\cite{Petekidis2004,Philippe2018} and numerical
simulations~\cite{Tsamados2010,Ikeda2012,Liu2016,Singh2020} have confirmed that glasses, the target of this letter, also exhibit HB-type rheological
behaviors.
Despite the ubiquity of the HB law, however, its structural origin
remains an open question even after more
than 90 years since the original paper by Herschel and Bulkley~\cite{Herschel1926}.

In this letter, we explore the structural signatures of the HB law in
sheared glasses by means of atomistic simulations.
We first show that the shear stress $\sigma$ suffers from finite size effects and thus that the HB exponent
cannot be determined by direct fitting.
To resolve this problem, we rely on the characteristic structural
information of plastic events, or avalanches,
which are responsible for the complex rheological response~\cite{Lemaitre2006}.
In particular, we numerically demonstrate that the instantaneous
normal modes
(INMs)~\cite{Bembenek1995,Stratt1995,Keyes1997,Gezelter1997,Bembenek2001}
allow us to extract structures of avalanches: so-called imaginary INMs
(Im-INMs), which have negative eigenvalues, correspond to shear
transformations (STs), the elementary processes of avalanches~\cite{Maloney2006}.
A phenomenological argument based on the criticality of yielding
transition further
enables us to determine the HB exponent $n$ from the shear rate
dependence of the statistics of Im-INMs.
The obtained value of $n$ is validated by the scaling collapse
of the shear stress.
With all these results, we conclude that Im-INMs serve as the
structural signatures of the HB law in sheared glasses.

\emph{System setup. ---}
We employ a two-dimensional ($d=2$) glass system introduced in
ref.~\cite{Oyama2020}.
The interparticle interaction is described by the Lennard-Jones potential with
smoothing terms that guarantee the smoothness of potential and force
at the cutoff distance $r^c_{ij}=1.3d_{ij}$, where $d_{ij}$ determines
the interaction range between particles $i$ and $j$.
To avoid crystallization, we consider a 50:50 binary mixture of
particles with an equal mass $m$ and different interaction ranges.
The interaction ranges for different combinations of particle types are
$d_{\rm SS}=5/6$, $d_{\rm
SL}=1.0$ and $d_{\rm LL}=7/6$, where subscripts $S$ and $L$
distinguish different species of particles.
The energy scale $\epsilon_{ij}=\epsilon = 1.0$ is set to be constant
regardless of particle species.
Throughout this letter, physical variables are nondimensionalized by the length unit
$d_{SL}$, the mass unit $m$ and the energy unit $\epsilon$.
We fix the number density to be
$\rho=N/L^2\sim 1.09$.
Initial configurations are obtained by minimizing the potential energy
of randomly generated particle distributions.
The thermal fluctuations are ignored.

We then apply external shear with different rates.
At each numerical step, purely affine shearing deformation of
strain $\Delta \gamma$ is applied
first, and then, the nonaffine dynamics are solved by integrating the
equations of motion under the Lees-Edwards boundary condition~\cite{Allen1987}.
To dissipate the input energy,
the drag force proportional to the
nonaffine velocity $\delta \boldsymbol{v}_i$ is exerted on each particle as $\boldsymbol{f}^{\rm
drag}_{
i}=-\Gamma \boldsymbol{v}_i$~\cite{Lemaitre2006,Salerno2012}.
In this work, we use
$\Gamma=1$.
To tune the shear rate, we fix the strain step at $\Delta
\gamma=1.0\times 10^{-7}$ and change the time step $\Delta t$.
By this protocol, plastic events are detected with the
same resolution regardless of the shear rate $\dot{\gamma}$.

\begin{figure}
\includegraphics[width=\linewidth]{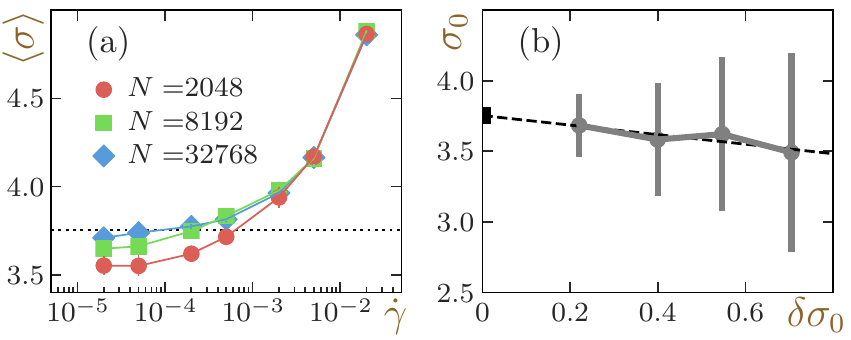}
\caption{
(a) Average steady-state stress $\langle{\sigma}\rangle$ as a function of the
shear rate $\dot{\gamma}$.
Different symbols stand
for different system sizes as shown in the legend.
Error bars represent the standard deviation between samples.
The dotted line marks $\sigma_{\rm Y}$ obtained in (b).
(b) Average steady-state stress under AQS shear ${\sigma}_0(L)$ at
different system sizes
as a function of their standard deviations $\delta\sigma_0(L)$.
The dashed line is the linear fit to the data, and the black square is
the estimated intrinsic yield stress $\sigma_{\rm
Y}\equiv{\sigma}_0(\infty)$ from the extrapolation of the data.
  \label{fig:flow_curves}}
\end{figure}

\emph{Flow curves of different system sizes. ---}
We start with the results of flow curves, or the plot of the stress $\sigma$ as a function
of the shear rate $\dot{\gamma}$, of different system sizes.
We prepared $N_s=8$ different samples for each system size and applied
simple shear at different rates in the range of $2\times
10^{-5}\le\dot{\gamma}\le 2\times 10^{-2}$.
For each sample and shear rate, we calculated the mean steady-state
stress $\bar{\sigma}$ from the data in the range of $1\le \gamma\le
4$, where $\gamma$ is the total amount of the applied strain.
In Fig.~\ref{fig:flow_curves}(a), we plot the sample-averaged steady-state stress $\langle {\sigma}\rangle\equiv \frac{1}{N_s}\sum_i^{N_s}\bar{\sigma}_i$ of different system sizes as functions of the shear rate $\dot{\gamma}$.
While the results of different system sizes match very well at high rates,
they exhibit discrepancies at low rates.
Because of these finite size effects, it is not clear which
part should be fitted, and we cannot determine the HB parameters,
namely, the yield stress $\sigma_{\rm Y}$ and the HB exponent $n$, by direct fitting.


The criticality of the yielding transition allows us to estimate
the yield stress $\sigma_{\rm Y}$, with the finite
size effects being properly taken into account.
First, given the criticality, it is natural to set
the following two scaling ansatzes:
$\xi\sim\Delta\sigma^{-\nu}$ and $\dot{\gamma}\sim\Delta\sigma^\beta$,
where $\Delta\sigma\equiv \langle \sigma\rangle-\sigma_{\rm Y}$ stands
for the distance to the critical point, $\nu$ and $\beta$ are critical exponents and
$\xi$ is the characteristic length.
In the current situation, $\xi$ corresponds to the average spatial expansion
of avalanches~\cite{Lin2014b}.
These avalanches are composed of multiple STs, or the elementary
processes of plastic events~\cite{Maloney2006}.

Now, let us consider a system under athermal quasistatic (AQS) shear, where the thermal fluctuations are absent and shear is imposed quasistatically.
Since the characteristic length reaches the system size ($\xi=L$) in
this situation,
utilizing the so-called statistical tilt symmetry~\cite{Fisher1998,Kolton2009,Lin2014b}, we can
express the system size dependence of
the steady-state average stress
${\sigma}_0(L)$ and
its fluctuations
$\delta\sigma_0(L)$
as ${\sigma}_0(L)=\sigma_{\rm Y}+k_1L^{-1/\nu}$ and
$\delta\sigma_0(L)=k_2L^{-1/\nu}$
respectively~\cite{Lin2014b},
where $k_1$ and $k_2$ are nonuniversal constants.
Because $\nu$ is positive by definition, the thermodynamic limit ($L\to \infty$) gives
the intrinsic yield stress $ \sigma_o(\infty)\equiv\sigma_{\rm Y}$,
which is accompanied by the disappearance of fluctuations $\delta\sigma_0(\infty)=0$.
We performed AQS simulations
whose details are found in ref.~\cite{Oyama2020} and measured
${\sigma}_0$
as a function of $\delta\sigma_0$, as shown in Fig.~\ref{fig:flow_curves}(b).
By extrapolating the data, we obtained the value of the intrinsic yield
stress {as
$\sigma_{\rm Y}\approx 3.753$}.
We note that the degree of system size
dependence depends on the details of systems, as
reflected by nonuniversal coefficients $k_1$ and $k_2$.
In fact, while some previous studies have reported system
size-dependent flow curves consistent with our results~\cite{Lemaitre2009,Chattoraj2010,Hatano2011,Roy2015},
such finite size effects were not observed in refs.~\cite{Tsamados2010,Singh2020}.

The statistical tilt symmetry further enables us
to derive
a scaling relation $n = 1-z/(d-d_f+z)$ that
provides the value of the HB exponent $n$, where $z$ is defined as $T\sim\xi^z$,
$T$ is the average time duration of
avalanches of size $\xi$, and $d_f$ is the fractal dimension of
avalanches~\cite{Kolton2009,Lin2014b}.
However, while $d_f$ can be measured by AQS simulations
as $d_f\approx 1.034$~\cite{Oyama2020},
$z$ cannot be measured in particle-based simulations in
principle.
Thus, we cannot utilize this scaling relation.
{To bypass this problem, we need another way to estimate $n$.
}

\begin{figure}
\includegraphics[width=\linewidth]{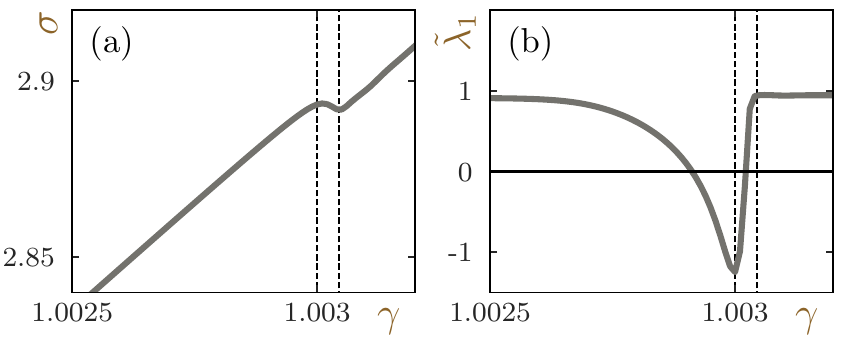}
\caption{
(a) Macroscopic stress $\sigma$ and (b) the lowest eigenvalue $\tilde{\lambda}_1$ as
functions of the applied strain $\gamma$.
Data are drawn from a system with $N=2048$ and
$\dot{\gamma}=2\times 10^{-5}$.
Dashed lines indicate the range during which a stress drop event is
taking place.
The event corresponding to the first peak in
Fig.~\ref{fig:INMs_sequence}(c) is shown.
}
  \label{fig:INM_stress}
\end{figure}

\begin{figure*}
\includegraphics[width=\linewidth]{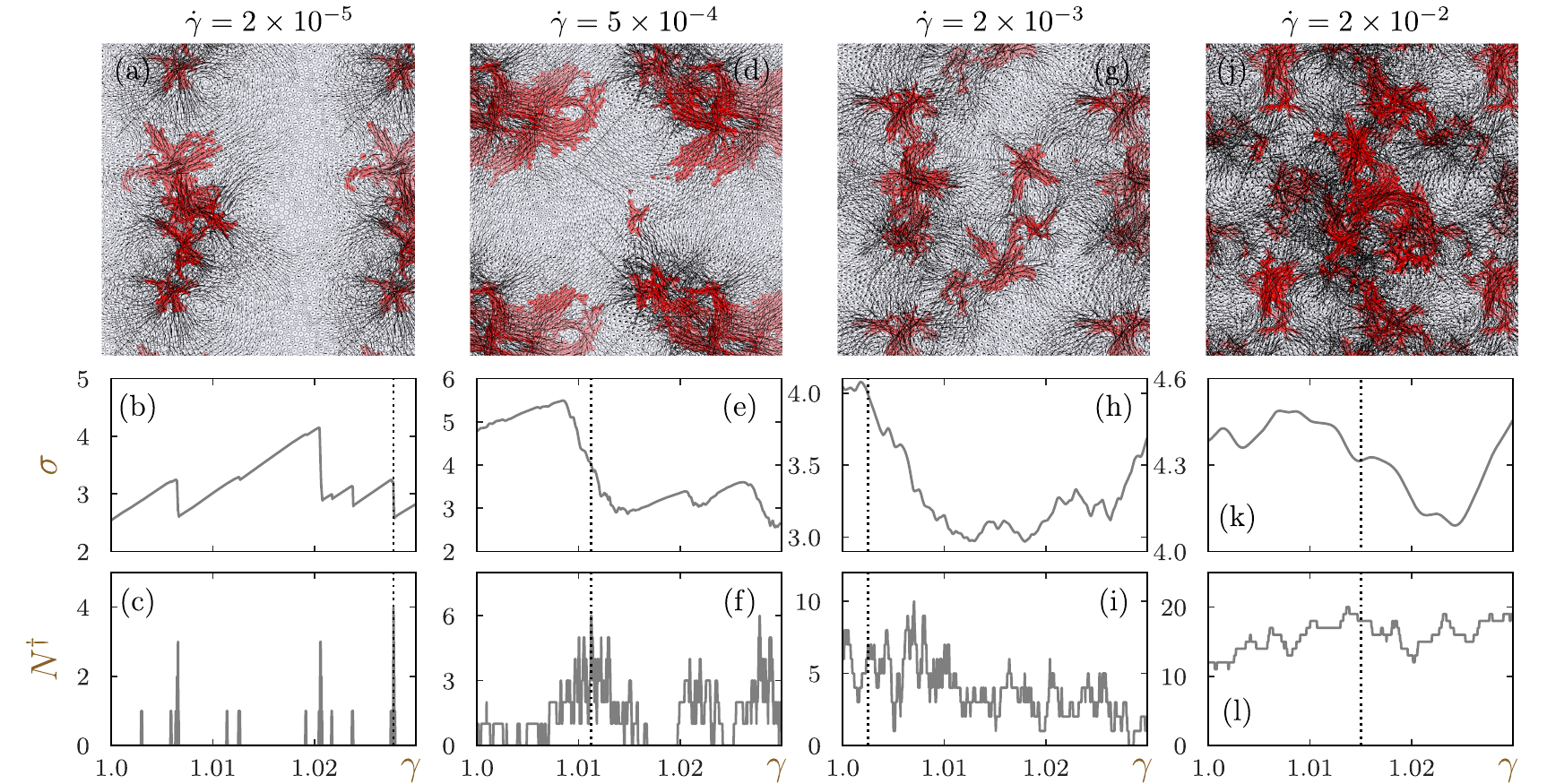}
\caption{
(Top row)
Visualization of Im-INMs at configurations indicated by dotted
lines in figures beneath.
All Im-INMs obtained for a given configuration are visualized on top of each other.
Black arrows depict eigenvectors, and only mobile particles (see the
Supplementary Material
SM1~\cite{Note2}
for the precise definition) in each
mode are highlighted in red.
The copied images due to the periodic boundary conditions are also
shown in lighter colors.
(Middle row)
The stress $\sigma$ and (bottom tow) the number of Im-INMs $N^\dagger$ as functions of the
applied shear $\gamma$.
Each column presents the results for different shear rates as indicated
above the top row.
All results are from the system with $N=2048$.
}
  \label{fig:INMs_sequence}
\end{figure*}

\emph{Instantaneous normal modes. ---}
We turn our attention to the structural signatures of
plastic events, which are responsible for the complex rheological
behaviors~\cite{Lemaitre2006}.
In the case of systems under AQS shear, normal mode (NM) analysis has revealed that the onset of plastic events
can be captured by the minimum eigenvalue of the dynamical matrix, or
the Hessian of the potential energy with respect to the particle
positions~\cite{Maloney2004}:
in the vicinity of the critical strain $\gamma_{\rm C}$ at which a
plastic event takes
place, the minimum eigenvalue $\lambda_1$ decreases as shear is
applied~\footnote{We ignore the trivial modes corresponding to the global translations.}.
$\lambda_1$ finally decays to zero at $\gamma_{\rm C}$ and excites a
corresponding ST, the elementary process of a plastic event.
The released energy from the excited ST propagates throughout the
system via the elastic field and can excite further secondary STs,
leading to an avalanche~\cite{Maloney2006}.
What if, then, we conduct similar analyses in systems under finite rate shear?
To tackle this simple question, we employed INM analysis
~\cite{Bembenek1995,Stratt1995,Keyes1997,Gezelter1997,Bembenek2001}.
The INMs are obtained as the eigenmodes of the dynamical matrix
calculated from the instantaneous particle configurations.
Under AQS shear,
the INMs are identical to the standard NMs
by definition.

We now present the results of the INM analysis under finite rate
shear.
We note that although the results for the system with $N=2048$ are explained in detail as a reference, the qualitative behaviors do not depend on the system sizes.
When only one single ST is excited under a very slow shear
($\dot{\gamma}=2\times 10^{-5}$),
we observe qualitatively similar behaviors to those of NMs under
AQS shear (Fig.~\ref{fig:INM_stress}):
all eigenvalues are positive in the elastic branch (where the
stress rises), and
the minimum eigenvalue $\tilde{\lambda}_1$ decreases drastically in the vicinity of the critical
strain $\gamma_{\rm C}$.
However, unlike the case of systems under AQS shear,
the stress does not start decreasing when $\tilde{\lambda}_1$ becomes zero.
Thus, $\tilde{\lambda}_1$ becomes even negative: the corresponding mode becomes
a so-called imaginary mode.
$\tilde{\lambda}_1$ then stops decreasing at $\gamma=\gamma_{\rm C}$, at
which the stress starts decreasing.
Since
the system requires nonzero time to dissipate the released energy from an excited
ST, the corresponding Im-INM survives over a finite strain even
beyond $\gamma_{\rm C}$.
{These results suggest that Im-INMs correspond to
evolving excited STs.}
In fact, as presented below, this speculation is the case even when
multiple Im-INMs are present simultaneously.

As stated above, STs sometimes form avalanches whose shapes
change drastically as $\dot{\gamma}$ increases.
To present the shear-rate-dependent change in structures of avalanches,
we further plot the stress $\sigma$ and the number of
Im-INMs $N^\dagger$ as functions of the applied strain $\gamma$
in Fig.~\ref{fig:INMs_sequence}.
When the rate is low ($\dot{\gamma}=2\times 10^{-5}$), peaks of $N^\dagger$ are
observed in a clearly synchronized manner with stress drop events (Fig.~\ref{fig:INMs_sequence}(b,c)).
In particular, large stress drop events are accompanied by multiple
Im-INMs.
If we visualize the obtained Im-INMs at a peak, we
find that STs corresponding to those modes form a quasilinear
avalanche (Fig.~\ref{fig:INMs_sequence}(a)).
When the shear rate becomes intermediate ($\dot{\gamma}=5\times
10^{-4}$), the amount of strain applied during the average lifetime of an ST becomes comparable to the typical strain interval between STs.
As a result, successive STs barely overlap temporally, as indicated by
a narrow peak with $N^\dagger=2$ {at approximately $\gamma=1.001$} (Fig.~\ref{fig:INMs_sequence}(e,f)).
Nevertheless, avalanche events, which are indicated by multiple Im-INMs and
take place less frequently, hardly overlap.
{The absence of overlaps of avalanches is manifested by
the fact that $N^\dagger$ becomes zero between avalanches.}
The visualization result (Fig.~\ref{fig:INMs_sequence}(d)) of the event
with the largest $N^\dagger$ is composed of a large system-spanning avalanche and an isolated ST.
When the shear rate becomes high ($\dot{\gamma}=2\times 10^{-3}$),
a strain large enough to induce secondary
avalanches is applied during the typical
lifetime of avalanches, and even avalanches start overlapping.
As a consequence, stress drop events and the peaks of $N^\dagger$
become obscured, and we cannot precisely locate them anymore (Fig.~\ref{fig:INMs_sequence}(h,i)).
The typical number of Im-INMs is notably larger than those under
slower shear, and
their visualization is composed of multiple
avalanches, as shown in Fig.~\ref{fig:INMs_sequence}(g).
The temporal overlap of avalanches is {the cause of the
decrease in the characteristic length of avalanches $\xi$ and
expected for the system with $\langle {\sigma}\rangle>\sigma_{\rm
Y}$~\cite{Lin2014b}.}
We stress that, indeed, $\langle {\sigma}\rangle > \sigma_{\rm Y}$
holds for $\dot{\gamma}\ge 2\times 10^{-3}$
(Fig.~\ref{fig:flow_curves}(a)).
If we increase the shear rate further ($\dot{\gamma}=2\times
10^{-2}$),
both stress drops and peaks of
$N^\dagger$ becomes completely obscured (Fig.~\ref{fig:INMs_sequence}(k,l)).
Although this result is presumably because many avalanches are simultaneously present, we can no longer decompose the visualized structure into individual avalanches (Fig.~\ref{fig:INMs_sequence}(j)).

\begin{figure}
\includegraphics[width=\linewidth]{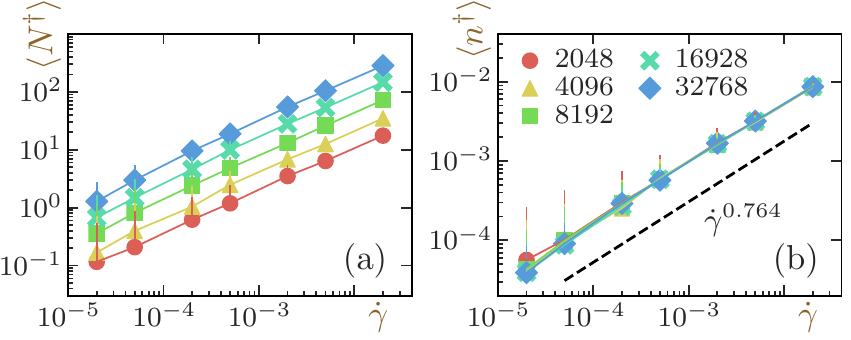}
\caption{
(a) Average number of Im-INMs $\langle N^\dagger\rangle$ and
(b) average Im-INM density $\langle n^\dagger\rangle$ as
functions of the shear rate $\dot{\gamma}$.
Different symbols are for
different system sizes as shown in the legend in (b).
The dashed line in (b) shows the power-law fitting result.
Averages are calculated over 250 randomly chosen independent configurations.} \label{fig:ndagger}
\end{figure}

To quantify these complicated shear-rate-dependent structural changes
of avalanches, we measure the average number of Im-INMs,
$\langle N^\dagger \rangle$, which carries information about the
typical number and size of avalanches.
We plot the results as functions of the shear rate $\dot{\gamma}$ in Fig.~\ref{fig:ndagger}(a).
As seen here, $\langle N^\dagger\rangle$ increases in a power-law manner regardless of the system size and the shear rate.
Furthermore, if we define the Im-INM density $\langle n^\dagger\rangle\equiv
\langle N^\dagger\rangle/N$, the results
for different system sizes all overlap without any parameters (Fig.~\ref{fig:ndagger}(b)).
These results suggest that $\langle N^\dagger\rangle$ can be expressed as $\langle
N^\dagger\rangle\sim N\dot{\gamma}^\lambda$, where the
exponent is estimated as $\lambda\sim 0.764$ by fitting
(All values of critical exponents are summarized in SM2~\footnote{See the Supplemental Material at [URL will be inserted by publisher]}).

\emph{{Scaling argument. ---}}
Now, we present an argument that reveals the relation
between the exponent $\lambda$ and the HB exponent $n$.
Above threshold ($\langle {\sigma}\rangle>\sigma_{\rm Y}$), there are multiple
avalanches with an average linear size of $\xi$.
By definition, the number of avalanches $N_{\rm ava}$ in this
situation is described as $N_{\rm ava}=(L/\xi)^d$~\cite{Lin2014b}.
Since avalanches possess $d_f$-dimensional fractal structures, we can
estimate the number of STs per avalanche $N_{\rm ST/ava}$ as $N_{\rm
ST/ava}\sim\xi^{d_f}$.
The expected average number of STs $\langle N^\dagger\rangle$ is then
expressed simply by the product of $N_{\rm ava}$ and $N_{\rm
ST/ava}$ as
\begin{align}
  \langle N^\dagger\rangle =N_{\rm ava}\times N_{\rm ST/ava}\sim
  L^d\cdot\xi^{d_f-d}\sim N\cdot\dot{\gamma}^n.\label{eq:1}
  \end{align}
Here, we utilized the relations $\nu=1/(d-d_f)$~\cite{Kolton2009,Lin2014b}, $n=1/\beta$
and $\xi^{-1/\nu}\sim \dot{\gamma}^{1/\beta}$ in the last equality.
This simple estimation accounts for the linear dependence of $N^\dagger$
on the system size $N$.
Furthermore, importantly, Eq.~\ref{eq:1} predicts that the exponent
$\lambda$
should coincide with the HB exponent $n$
\footnote{
The simple estimation here itself is applicable only to the situations
with \unexpanded{$\langle\sigma\rangle > \sigma_{\rm Y}$}, which is met for large systems and the
high shear rate regime, although the same power-law behavior is
observed for all system sizes and shear rates.}.

\begin{figure}
\includegraphics[width=\linewidth]{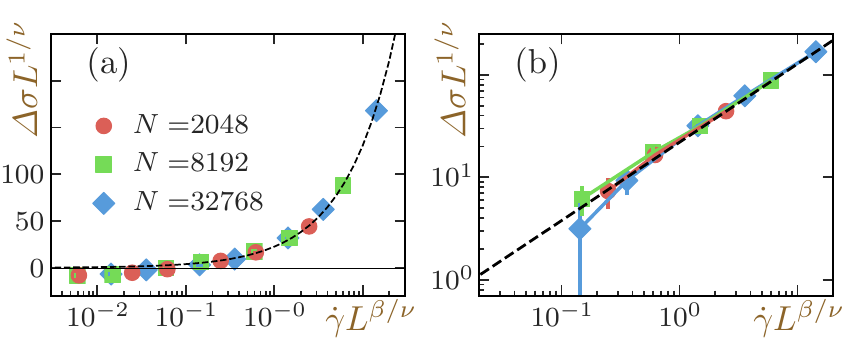}
\caption{
Distance to the critical point $\Delta \sigma$ as a function of
the shear rate $\dot{\gamma}$ with finite size scaling.
(a) Semi-log plot. (b) Log-log plot.
Different symbols express different system sizes, as shown in the legend.
The dashed line represents the HB law with  parameters estimated in
Figs.~\ref{fig:flow_curves} and \ref{fig:ndagger} and Eq.~\ref{eq:1}.
}  \label{fig:HBs4}
\end{figure}

To further check the reliability of the obtained value of the HB
exponent $n=\lambda \approx 0.764$,
now, we consider the finite size scaling of the flow curves.
The scaling ansatzes simply imply the scaling relation $\dot{\gamma}\sim L^{-\beta/\nu}f(\Delta\sigma
L^{1/\nu})$, where $f(x)$ is a suitable scaling function~\cite{Lin2014b}.
As shown in Fig.~\ref{fig:HBs4}(a), the results for different system sizes
are collapsed very well with this relation.
Furthermore, the curve of the HB law precisely captures the simulation
data meeting $\Delta\sigma > 0$ (Fig.~\ref{fig:HBs4}(b)).
Note that the exponents $n$ and $\beta$ meet the relation $n=1/\beta$.
{These results mean that the HB exponent can be accurately
estimated by the average number of Im-INMs,
and thus, we conclude that these modes serve as the structural
signatures of the HB law.}

{We stress that mean-field theory~\cite{Hebraud1998} predicts the value of $n=0.5$, and several
studies have reported consistent values~\cite{Lemaitre2009,Dinkgreve2015a,Vasisht2018,Cabriolu2019,Singh2020}.
We show in SM3~\cite{Note2} that if we ignore
the finite size effects and fit to the whole data of a single system
size directly as in previous studies, we also obtain consistent values $n\approx 0.5$.}

\emph{Summary and overview. ---}
To summarize,
we first showed that Im-INMs
provide information about evolving STs.
Then, by investigating the shear rate-dependent development of
complicated structures of avalanches,
we further showed that the average number
of Im-INMs exhibits a power-law dependence on the shear rate
with the same exponent as the HB law.
These findings were further validated by the success of scaling
collapse of flow curves with different system sizes.
From all these results, we conclude that Im-INMs are
the structural signatures of the HB rheology of sheared glasses.

It would be very important to investigate whether the findings in this letter
are applicable
to other soft materials that obey the HB law, such as
suspensions or emulsions, where the jamming criticality also plays a major
role.
Investigating the effects of the introduction of thermal~\cite{Karmakar2010,Chattoraj2010,Chattoraj2011} or various types of active
noises~\cite{Fily2012,Berthier2017,Oyama2019b} would also provide
useful knowledge for material design.

\begin{acknowledgments}
{
This work was financially supported by KAKENHI grants
(nos. 18H05225, 19H01812, 19K14670, 20H01868, 20H00128, 20K14436 and 20J00802) and partially supported by the Asahi Glass Foundation.
}
\end{acknowledgments}

%
%

%
%
\end{document}